\documentclass[preprint,showkeys,preprintnumbers,amsmath,amssymb]{revtex4}

\usepackage{graphicx}
\usepackage{dcolumn}
\usepackage{bm}
\usepackage{color}

\newcommand{\ff}{\frac{1}{2}}

\begin{document}

\title{New insights on frequency combinations and 'forbidden frequencies' in de Haas-van Alphen spectrum of $\kappa$-(ET)$_2$Cu(SCN)$_2$}

\author{Alain Audouard${^1}$, Jean-Yves~Fortin$^{2}$, David Vignolles$^1$, Vladimir~N.~Laukhin$^{3,4}$, Nataliya~D.~Kushch$^5$ and  Eduard B. Yagubskii$^5$}

\affiliation{$^1$ Laboratoire National des Champs Magn\'{e}tiques
Intenses (UPR 3228 CNRS, INSA, UGA, UPS) 143 avenue de Rangueil,
F-31400 Toulouse, France.\\$^2$ Institut Jean Lamour, D\'epartement de Physique de la
Mati\`ere et des Mat\'eriaux,
CNRS-UMR 7198, Vandoeuvre-les-Nancy, F-54506, France\\$^3$ Instituci\'{o} Catalana de Recerca i
Estudis Avan\c{c}ats (ICREA), 08210 Barcelona, Spain.\\$^4$
Institut de Ci\`{e}ncia de Materials de Barcelona, Consejo
Superior de Investigationes Cient\'{i}ficas, Campus Universitat
Aut\`{o}noma de Barcelona, Bellaterra 08193, Spain.\\$^5$
Institute of Problems of Chemical Physics, Russian
Academy of Sciences, 142432 Chernogolovka, MD, Russia.}%

\date{\today}

\begin{abstract}{De Haas-van Alphen oscillations of the organic metal
$\kappa$-(ET)$_2$Cu(SCN)$_2$ have been measured up to 55 T at liquid helium
temperatures. The Fermi surface of this charge transfer salt is a textbook
example of linear chain of orbits coupled by magnetic breakdown. Accordingly,
the oscillation spectrum is composed of linear combinations of the frequencies
linked to the $\alpha$ and magnetic breakdown-induced $\beta$ orbits. The field
and temperature dependence of all the observed Fourier components, in particular the 'forbidden frequency' $\beta-\alpha$ which cannot correspond to a classical orbit,  are quantitatively accounted for by analytical calculations based on a second order
development of the free energy, i.e. beyond the first order
Lifshitz-Kosevich formula.
}
%

\end{abstract}
\maketitle

%
\section{Introduction}
\label{sec:intro}

The family of organic charge transfer salts  $\kappa$-(ET)$_2X$, where ET stands for the bis(ethylenedithio)tetrathiafulvalene molecule and $X$ is a monovalent cation,  arouses interest for many years. Indeed, its members can be classified in a very rich phase diagram including ground states ranging from Mott insulator to superconductor \cite{Ka06,Ar12}. $\kappa$-(ET)$_2$Cu(SCN)$_2$, which has been synthesized as early as 1988 \cite{Ur88}, is on the metallic side of this phase diagram and is still one of the most studied organic superconductor. Illustrating this long-standing interest, the existence of a Fulde-Ferrell-Larkin-Ovchinnikov inhomogeneous superconducting state in this salt, considered more than 20 years ago for low dimensional conductors \cite{Sh94,Du95}, have received first experimental hints in the early 2000s \cite{Si00}  and have been recently confirmed \cite{Ma14}.

Remarkably, the Fermi surface of $\kappa$-(ET)$_2$Cu(SCN)$_2$
\cite{Os88,Ju89,Ca96} (see Fig.~\ref{d_TF}) is the first experimental
realization of the model Fermi surface proposed by Pippard in the early sixties
to compute the effect of magnetic breakdown (MB) in multiband metals
\cite{Pi62}. Both Shubnikov-de Haas (SdH)
\cite{To88,Mu90,Sa90,An91,He91,Ca94,Ha96,Ka96} and de Haas-van Alphen (dHvA)
\cite{Wo92,Me95,Uj97,St99,Gv02} oscillations have been widely studied in this
compound. Provided the magnetic field strength is high enough to overcome the MB
gap between the $\alpha$ orbit and the quasi one-dimensional sheets, linear
combinations of the frequencies linked to the  $\alpha$  and MB-induced $\beta$
orbits are observed in the oscillatory spectrum (for a review see e.g.
\cite{Uj08}), as expected in the framework of the coupled orbits network model
of Falicov-Stachowiak \cite{Fa66}. Within this model, the frequency $F_{\eta}$
of any Fourier component can be expressed  as  $F_{\eta}$ =
$n_{\alpha}F_{\alpha}+n_{\beta}F_{\beta}$ where $n_{\alpha(\beta)}$ is an
integer and $F_{\alpha(\beta)}$ is the frequency linked to the $\alpha(\beta)$
orbit. With the exception of quantum interference (QI), which only holds for
magnetoresistance, and for which the effective mass is the difference between
the partial effective mass of each QI path \cite{St71}, the effective mass
$m_{\eta}$ is given by $m_{\eta}$ = $n_{\alpha}m_{\alpha}+n_{\beta}m_{\beta}$.
However, limiting ourselves to dHvA data, the field and temperature dependence
of few of the observed Fourier components does not behave consistently with
this model. For example, the effective masses linked to $2\alpha$ and $2\beta$,
determined in the framework of the Lifshitz-Kosevich model \cite{Sh84}, are
strongly different from 2$m_{\alpha}$ and 2$m_{\beta}$, respectively
\cite{Uj97}. In addition, puzzling data are reported regarding the component
$\beta$ + $\alpha$ which corresponds to a MB orbit. Indeed, even though
according to \cite{Me95} $m_{\beta+\alpha}$ is equal to $m_{\beta}$+$m_{\alpha}$
within the error bars, in agreement with the Falicov-Stachowiak model, this
relation cannot account for the data of \cite{Uj97,St99} for which
$m_{\beta+\alpha}$  is definitely smaller than the sum $m_{\beta}+m_{\alpha}$
(see \cite{Au14} for a discussion of this feature). In addition to all this, the
most striking discrepancy between the Falicov-Stachowiak model and the dHvA data
(for which QI is not relevant since dHvA is only sensitive to the density of
states) is the presence of the Fourier components $\beta-\alpha$ and
2($\beta-\alpha$), known as `forbidden frequencies' which cannot correspond to
closed orbits (for a review, see \cite{Ka04,Uj08}).

These features have been addressed in the case of $\theta$-phase salts belonging
to the family $\theta$-(ET)$_4$MBr$_4$(C$_6$H$_4$Cl$_2$) for M = Co, Zn
\cite{Au12,Au14,Au15} which share with $\kappa$-(ET)$_2$Cu(SCN)$_2$ the same
Fermi surface topology. Briefly, the strong two-dimensionality of these
compounds is responsible for a significant oscillation of the chemical potential
in magnetic field which requires a calculation of the free energy (canonical
ensemble) up to the second-order in damping factors. In few cases, the first
order term of the oscillatory magnetization, which corresponds to the usual
Lifshitz-Kosevich formula \cite{Sh84}, may have a smaller contribution to the
amplitude  of the Fourier components linked to MB orbits than the second order
term. Moreover, the second order term solely accounts for the `forbidden
frequencies'. A very good quantitative agreement between these calculations and
the reported dHvA data is observed. Nevertheless, as pointed out in \cite{Au14},
data analysis relevant to other salts, with different physical parameters, such
as effective masses, MB field, $etc.$, are required. Therefore, the aim of this
paper is to report on dHvA data of $\kappa$-(ET)$_2$Cu(SCN)$_2$ at high magnetic
field. It is confirmed that the usual Lifshitz-Kosevich formalism cannot account
for few of the Fourier component  amplitudes such as 2$\alpha$ and, obviously,
$\beta-\alpha$. In contrast, it is demonstrated that the above mentioned second
order development holds for all the observed Fourier components.

\section{Experimental}

The studied crystal, with approximate dimensions 0.1 $\times$ 0.1 $\times$ 0.04 mm$^3$, was synthesized by
electrocrystallization technique. Magnetic torque measurements were performed in pulsed magnetic fields of up to  55 T with pulse
decay duration of 0.36 s, in the temperature range from 1.7 K to
4.2 K. The angle between the magnetic field direction and the normal to the conducting plane was $\theta$ = 20$^{\circ}$. Fig.~\ref{d_TF}(a) displays magnetic torque data at 1.7 K. As expected, a strong hysteresis is observed in the superconducting state at low field due to flow of magnetic flux either into or out of the crystal as the field either increases or decreases \cite{Mo01}. In contrast, data for up and down fields are indiscernible from each other in the range where oscillations are observed. This result indicates that any temperature change due to eventual eddy currents is negligible. In addition, torque interaction effect might have been considered since, due to magnetic torque,  $\theta$ angle is liable to change during the field sweep for small cantilever stiffness and (or) high $\theta$ value \cite{Be99}. In the present case, $\theta$ changes are always smaller than 0.1$^\circ$, leading to negligibly small amplitude and frequency changes of at most 0.2 $\%$ at the highest fields explored.  In the following, only data recorded during the decreasing part of the field is therefore considered.
Analysis of the oscillatory part of the magnetic torque is based on
discrete Fourier transforms of the data, after
substraction of a smoothly varying background, calculated with a
Blackman window.

\section{\label{sec:results}Results and discussion}

\begin{figure}[h!]                                                    
\centering
\resizebox{0.6\columnwidth}{!}{\includegraphics*{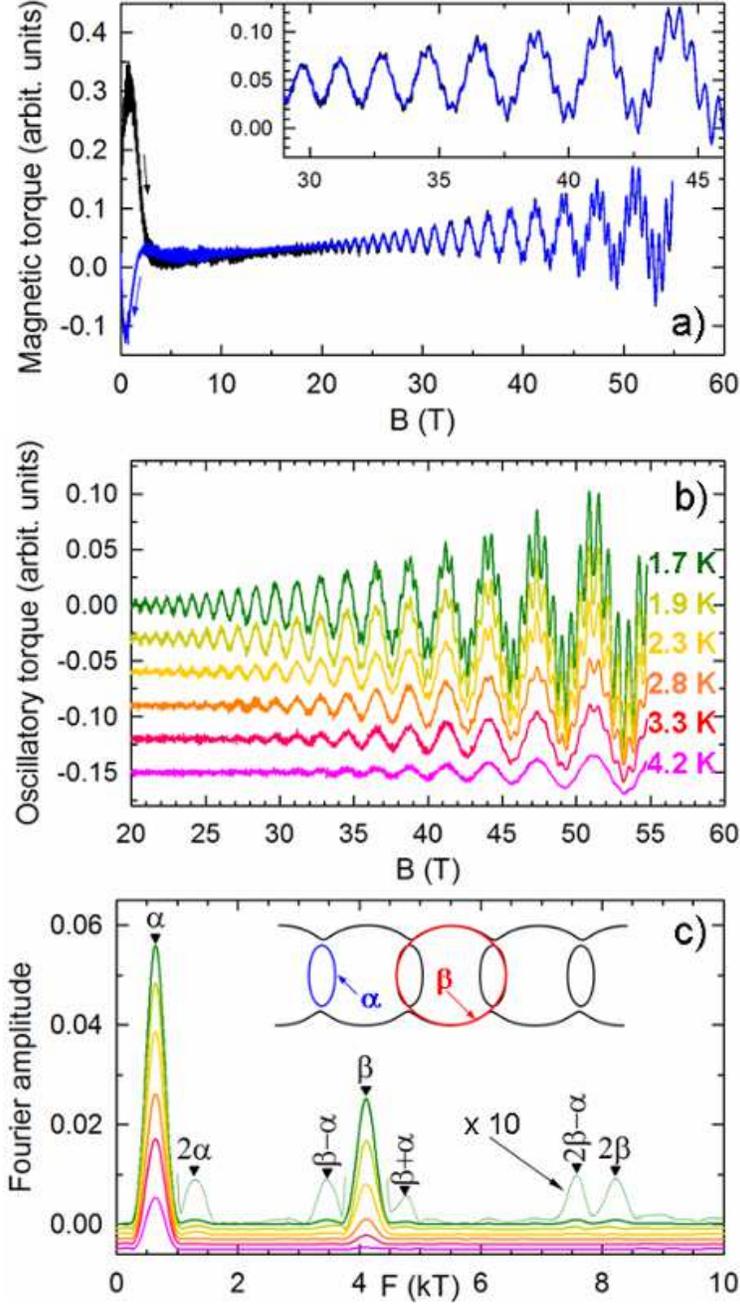}}
\caption{\label{d_TF} (a) Field-dependent magnetic torque at T = 1.7 K. Black and blue lines stand for increasing and decreasing field sweeps, respectively. (b) Field-dependent oscillatory torque for decreasing field sweeps at various temperatures and (c) corresponding Fourier analysis in the field range 20-55 T. The angle between the magnetic field direction and the normal to the conducting plane is $\theta$ = 20$^{\circ}$. Marks are calculated with  $F_{\alpha} (\theta=0)$ = 0.600 kT and
$F_{\beta} (\theta=0)$ = 3.861 kT. The insert displays a sketch of the Fermi surface in which the $\alpha$ and $\beta$ orbits are indicated.}
\end{figure}

The high field range of the oscillatory part of the torque data is displayed in
Fig.~\ref{d_TF}(a). As expected, corresponding Fourier analysis reveals several
components, of which the frequencies are linear combinations of  $F_{\alpha}$
and  $F_{\beta}$. Owing to the tilt angle $\theta$= 20$^{\circ}$, these frequencies are
$F_{\alpha}(\theta=0)$ = 0.600(1) kT and $F_{\beta}(\theta=0)$ = 3.861(7) kT.
These values are in good agreement with the data of the literature
\cite{To88,Mu90,Sa90,An91,He91,Ca94,Ha96,Ka96,Wo92,Me95,Uj97,St99,Gv02} in which
values in the range $F_{\alpha}(\theta=0)$ = 0.596(2)-0.670(4) kT  and
$F_{\beta}(\theta=0)$ = 3.8-3.923(9) kT are reported. The corresponding
$\alpha$ orbit area amounts to 16 \% of the first Brillouin zone area, in
agreement with both previous experimental data and band structure calculations
\cite{Os88,Ju89,Ca96}.

\begin{figure}                                                    
\centering
\resizebox{0.75\columnwidth}{!}{\includegraphics*{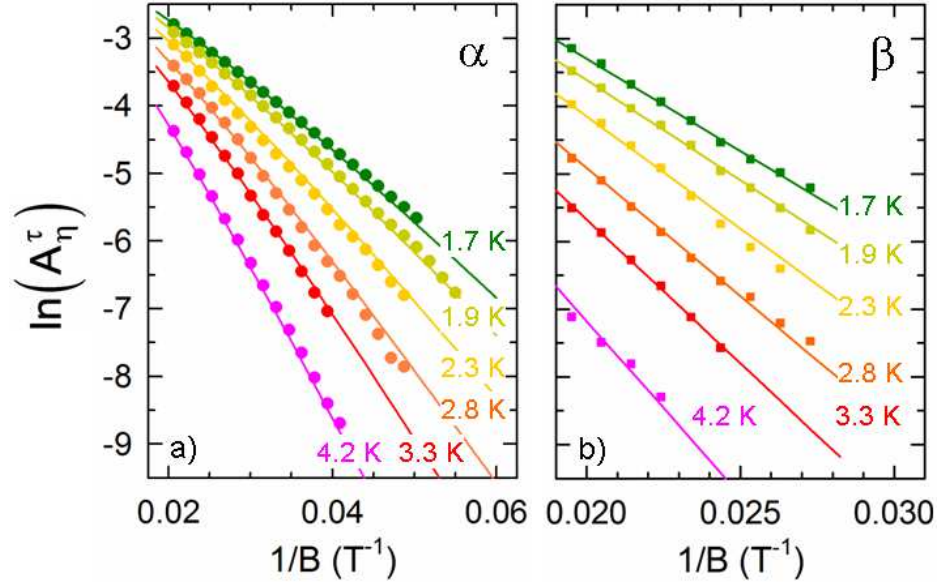}}
\caption{\label{Dingle_alpha_beta} Dingle plots of (a) $\alpha$ and (b) $\beta$ Fourier components. Symbols are experimental Fourier amplitudes and solid lines in (a) and (b) are best fits of Eqs.~\ref{Eq:alpha} and~\ref{Eq:beta} , respectively, obtained with $m_{\alpha}$ = 3.07 , $m_{\beta}$ = 6.03 , $T_D$ = 0.87 K and $B_0$ = 16 T. Uncertainty on these parameters is given in the text.}
\end{figure}

According to the calculations reported in Refs.~\cite{Au12,Au14,Au15}, the
amplitude of the basic $\alpha$ and $\beta$ Fourier components, up to second
order in damping factors, is accounted for by:

\begin{equation}
\label{Eq:alpha}
A_{\alpha}=-\frac{F_{\alpha}}{\pi m_{\alpha}}R_{\alpha,1}-
\frac{F_{\alpha}}{\pi m_{\beta}}
\left [
\ff R_{\alpha,1}R_{\alpha,2}
+\frac{1}{6}R_{\alpha,2}R_{\alpha,3}+2R_{\beta,1}R_{\beta+\alpha,1}
+\ff R_{\beta,2}R_{2\beta-\alpha,1}
\right ]
\end{equation}

and

\begin{equation}
\label{Eq:beta}
A_{\beta}=-\frac{F_{\beta}}{\pi m_{\beta}}R_{\beta,1}-
\frac{F_{\beta}}{\pi m_{\beta}}
\left [
\ff R_{\beta,1}R_{\beta,2}
+\frac{1}{6}R_{\beta,2}R_{\beta,3}+2R_{\alpha,1}R_{\beta+\alpha,1}
+2R_{\beta,1}R_{2\beta,1}
\right ],
\end{equation}

respectively. Damping factors can be expressed as $R_{\eta,p}(B,T)$ = $R^T_{\eta,p}(B,T) R^{D}_{\eta,p}(B) R^{MB}_{\eta,p}(B) R^{s}_{\eta,p}$,
where $\eta$ stands either for $\alpha$ or $\beta$ and $p$ is the harmonic
number. For a two-dimensional Fermi surface, the thermal, Dingle, MB and spin
damping factors are expressed as $R^{T}_{\eta,p}= pu_{\eta}
\sinh^{-1}(pu_{\eta})$, $R^{D}_{\eta,p} =
\exp(-pu_0m_{\eta}T_{D\eta}/B\cos \theta)$, $R^{MB}_{\eta,p} =
(ip_0)^{n^t_{\eta}}(q_0)^{n^r_{\eta}}$ and $R^{s}_{\eta,p} = \cos(\pi
g^*_{\eta}m_{\eta}/2\cos\theta)$, respectively \cite{Sh84}. The parameter
($u_{\eta}$) and the constant ($u_0$) are expressed as $u_{\eta}$ = $u_0
m_{\eta} T/B\cos\theta$ and $u_0$ = 2$\pi^2 k_B m_e(e\hbar)^{-1}$ = 14.694
T/K. $T_{D\eta}$ is the Dingle temperature
defined by $T_{D}$ = $\hbar(2\pi k_B\tau)^{-1}$, where $\tau^{-1}$ is the
scattering rate. $m_{\eta}$ and $g_{\eta}$ are the
effective mass (in $m_e$ units) and effective Land\'{e} factor, respectively.
The  MB damping factor involves the number of tunnelings ($n^t_{\eta}$) and
reflections ($n^r_{\eta}$) encountered by a quasiparticle traveling the
orbit. The corresponding probabilities, $p_0$ and $q_0$=$(1-p_0^2)^{1/2}$,
respectively, are given by the Chambers formula  which involve the magnetic
breakdown field $B_0$: $p_0^2$ = $e^{-B_0/B\cos\theta}$ \cite{Ch66}.

The first order term of these equations just corresponds to the
Lifshitz-Kosevich formula. Besides, provided that the spin damping factors
$R^{s}_{\alpha(\beta)}$ are not too small (i.e. $m_{\eta}g_{\eta}/\cos\theta)$
is not close to an odd integer), the second order term is negligible
\cite{Au14}. This point, which can be checked $a~posteriori$, allows us to
perform the data analysis relevant to $\alpha$ and $\beta$ in the framework of
Lifshitz-Kosevich formula.  Nevertheless, even though the spin damping factors
act as constant prefactors in this case, 5 parameters are still involved,
namely $m_{\alpha}$, $m_{\beta}$, $T_{D\alpha}$, $T_{D\beta}$ and $B_0$. For
this reason, it is assumed in the following that the Dingle temperature is the
same for the $\alpha$ and $\beta$ orbits ($T_{D\alpha}$=$T_{D\beta}$=$T_D$).
Finally, it must be kept in mind that the magnetic torque amplitudes
$A^{\tau}_{\eta}$ are related to the dHvA amplitude by $A^{\tau}_{\eta}$ =
$\tau_0BA_{\eta}$ where $\tau_0$ is a prefactor depending on the crystal
mass, cantilever stiffness and tilt angle $\theta$.

Dingle plots relevant to the $\alpha$ and $\beta$ orbits are displayed in Fig.~\ref{Dingle_alpha_beta}. A good agreement between the data and the calculations is obtained with $m_{\alpha}$ = 3.07(4), $m_{\beta}$ = 6.03(11), $T_D$ = 0.87(6) K and $B_0$ = 16(8) T. The effective mass values are within the range of the data reported in the literature in which $m_{\alpha}$ and $m_{\beta}$ ranging from 3 to 3.6  and 5.6 to 7.1, respectively, can be found. In line with statement of Ref. \cite{Me97}, $m_{\beta}$/$m_{\alpha}$ = 2 within the error bars.

It can be noticed that the MB field $B_0$ is obtained with a large uncertainty. This feature, can be explained considering that, in addition to an increase of the Dingle plot slope relevant to the $\beta$ orbit, the main discernible effect of MB is to induce a curvature in the field dependence of the $\alpha$ amplitude. However, this curvature remains small in the explored field range which yields a large uncertainty. This uncertainty is reflected in the strong scattering of the previously reported experimental data which yield $B_0$ values (themselves with large uncertainties) in the range from 15 T to 45 T \cite{Ca94,Me95,Ha96,Uj97,Gv02,Mi06}.

\begin{figure*}                                                    
\centering
\resizebox{0.8\columnwidth}{!}{\includegraphics*{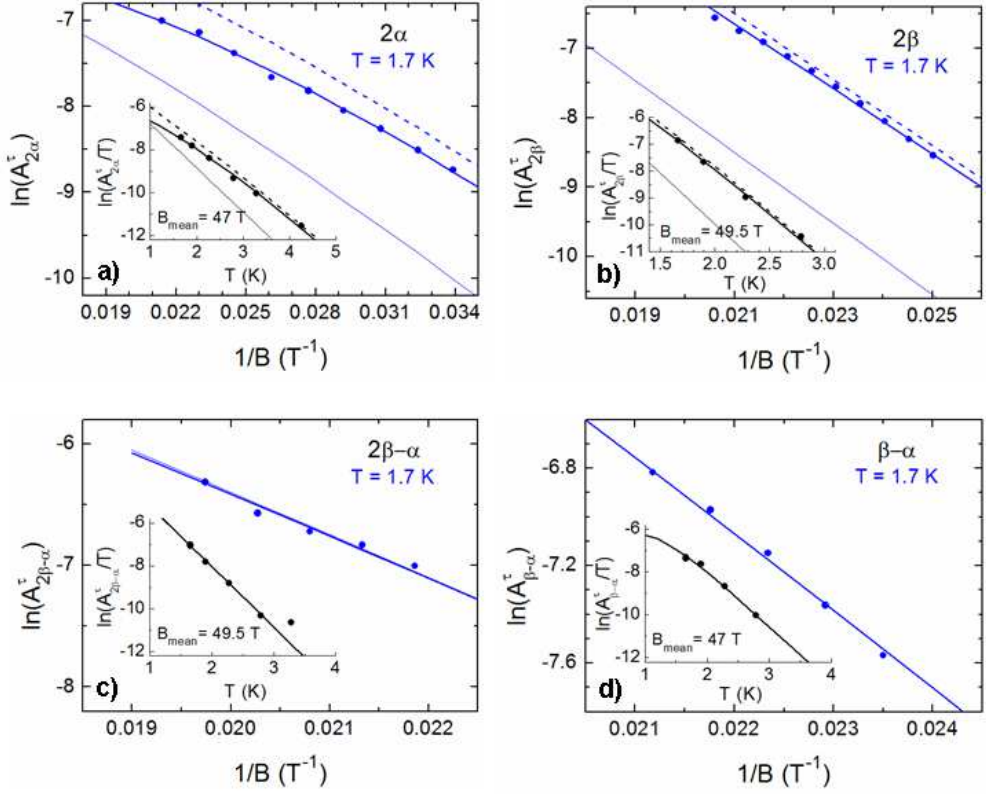}}
\caption{\label{Dingle_mass-plot_combinaisons} Dingle plots of (a)  $2\alpha$,
(b)  $2\beta$, (c)  $2\beta-\alpha$ and (d) $\beta-\alpha$ Fourier components.
Mass plots are reported in the inserts. Symbols are experimental Fourier
amplitudes and solid lines are best fits of Eqs.~\ref{Eq:2alpha},
\ref{Eq:2beta}, \ref{Eq:2beta-alpha} and \ref{Eq:beta-alpha}, respectively, to
the data, obtained with the same parameters as in Fig.~\ref{Dingle_alpha_beta}
and with $g^*_{\alpha}$ = 1.73  and $g^*_{\beta}$ = 1.52. Uncertainty on these parameters is given in the text. Thin solid and dashed
lines are the contribution of the first (i.e. Lifshitz-Kosevich) and second
order terms, respectively, of the above mentioned equations. }
\end{figure*}

Data relevant to frequency combinations are reported in Fig.~\ref{Dingle_mass-plot_combinaisons}. Let us consider first the component 2$\alpha$, the amplitude of which is given by

\begin{equation}
\label{Eq:2alpha}
A_{2\alpha}=-\frac{F_{\alpha}}{2\pi m_{\alpha}}R_{\alpha,2}+
\frac{F_{\alpha}}{\pi m_{\beta}}\left [
R_{\alpha,1}^2-\frac{2}{3}R_{\alpha,1}R_{\alpha,3}-R_{\beta,2}
R_{\beta+\alpha,2} \right ].
\end{equation}

The first order term  of Eq.~\ref{Eq:2alpha}, involving $R_{\alpha,2}$,
corresponds to the Lifshitz formula. The second order term is dominated by the
factor involving $R_{\alpha,1}^2$. Putting aside the spin damping factors,
$R_{\alpha,1}^2$ is strictly equal to $R_{\alpha,2}$ as the ratio $B/T$ goes to
zero. In the explored high field range, $R_{\alpha,1}^2$ takes even larger
values than $R_{\alpha,2}$. As a consequence, both the first and second order
terms contribute to the amplitude. Hence, the two involved spin damping
factors, $R^s_{\alpha,2}$ and $R^s_{\alpha,1}$, must be taken into account.
Solid lines in Fig.~\ref{Dingle_mass-plot_combinaisons}(a) are best fits of
Eq.~\ref{Eq:2alpha}. They are obtained with the same effective masses, Dingle
temperature and magnetic breakdown field as for the data relevant to $\alpha$
and $\beta$ (see Fig.~\ref{Dingle_alpha_beta}) and with $g_{\alpha}$ = 1.73(5).
This latter value is slightly larger than deduced from the angle dependence of
the $\alpha$ oscillations ($g_{\alpha}$ = 1.63(8) \cite{Me95}). Nevertheless, it
can be remarked that the product $g_{\alpha}m_{\alpha}$ = 5.6(2), governing the
spin damping factor is close to the  value,  $g_{\alpha}m_{\alpha}$ = 5.2(5),
obtained in Ref.~\cite{Me95}. Strikingly, it can be observed that the second
order term has a dominant contribution to the amplitude of the component
2$\alpha$.

Similar considerations hold for 2$\beta$. Indeed, its amplitude is accounted for
by

\begin{equation}
\label{Eq:2beta}
A_{2\beta}=-\frac{F_{\beta}}{2\pi m_{\beta}}\left
[R_{\beta,2}+2R_{2\beta,1}\right ]+
\frac{F_{\beta}}{\pi m_{\beta}}
\left [
R_{\beta,1}^2-\frac{2}{3}R_{\beta,1}R_{\beta,3}-\frac{1}{4}R_{\beta,2}R_{\beta,4
}-R_{\alpha,2}R_{\beta+\alpha,2}\right.
\end{equation}
\begin{equation}
\nonumber
+\left .2R_{\alpha,1}R_{2\beta-\alpha,1}-R_{\beta,2}R_{
2\beta,2}-R_{\beta,4}R_{2\beta,1}
\right. ].
\end{equation}

The Lifshitz-Kosevich contribution arises from both the magnetic breakdown orbit
2$\beta$ involving 4 tunneling and 2 reflections and the second harmonics of
$\beta$ governed by $R_{2\beta,1}$ and $R_{\beta,2}$, respectively (see
\cite{Au14} for a discussion of this point). The factors $R_{\beta,1}^2$, and to
a less extent $R_{\alpha,1}R_{2\beta-\alpha,1}$, appearing in Eq.~\ref{Eq:2beta}
play a role similar to  $R_{\alpha,1}^2$ in Eq.~\ref{Eq:2alpha}, the other
second order factors being negligible. Solid lines in
Fig.~\ref{Dingle_mass-plot_combinaisons}(b) are best fits of Eq.~\ref{Eq:2beta}.
They are obtained with the same set of parameters as in
Fig.~\ref{Dingle_alpha_beta} and  $g_{\beta}$ = 1.52(12) which is just the value
given in Ref.~\cite{Gv02}. Similarly to 2$\alpha$, the second order term
dominates the amplitude of 2$\beta$.

The amplitude of the component $2\beta-\alpha$, which corresponds to a classical magnetic breakdown orbit involving 4 tunneling and 2 reflections, is given by

\begin{equation}
\label{Eq:2beta-alpha}
A_{2\beta-\alpha}=-\frac{F_{2\beta-\alpha}}{\pi m_{2\beta-\alpha}}
R_{2\beta-\alpha,1}-
\frac{F_{2\beta-\alpha}}{\pi m_{\beta}}
\left [\ff R_{\alpha,1}R_{\beta,2}+
\frac{1}{3}
R_{\alpha,3}R_{\beta+\alpha,2}\right ].
\end{equation}

The second order term is dominated by the product $R_{\alpha,1}R_{\beta,2}$
which is much lower than the damping factor $R_{2\beta-\alpha,1}$ entering the
Lifshitz-Kosevich formula, provided the spin damping factor
$R^s_{2\beta-\alpha}$ is not too small. For this reason, the Lifshitz-Kosevich
formula holds for this orbit as it is the case for the above considered $\alpha$
and $\beta$ orbits and accounts for the data of
Fig.~\ref{Dingle_mass-plot_combinaisons}(c).

Finally, the `forbidden orbit' $\beta-\alpha$ is accounted for by

\begin{equation}
\label{Eq:beta-alpha}
A_{\beta-\alpha}=-\frac{F_{\beta-\alpha}}{\pi m_{\beta}}\left [
R_{\alpha,1}R_{\beta,1}+R_{\alpha,2}R_{\beta+\alpha,1}+R_{\beta,2}R_{
\beta+\alpha,1}+R_{\beta,1}R_{2\beta-\alpha,1}\right ].
\end{equation}

Since this component does not correspond to a classical orbit, only a second
order term, dominated by the product $R_{\alpha,1}R_{\beta,1}$, enters the
amplitude. Solid line in Fig.~\ref{Dingle_mass-plot_combinaisons}(d) are best
fits to Eq.~\ref{Eq:beta-alpha}. They are obtained with the same set of
parameters as in Fig.~\ref{Dingle_alpha_beta}, as well. A very good agreement is
observed.

\section{\label{sec:Conclusion}Summary and conclusion}

De Haas-van Alphen oscillations of the quasi-two dimensional organic metal
$\kappa$-(ET)$_2$Cu(SCN)$_2$ have been studied at high magnetic field. The Fermi
surface of this emblematic organic metal is a prime example of a linear chain of
orbits coupled by magnetic breakdown. Accordingly, the oscillation spectrum is
composed of linear combinations of the frequencies linked to the $\alpha$ and
magnetic breakdown-induced $\beta$ orbits. The field and temperature dependence
of the observed Fourier components is consistently analyzed in the
framework of analytical calculations based on a second order development of the
free energy taking into account the oscillations of the chemical potential,
i.e. beyond the first order Lifshitz-Kosevich formula, and with the same set of
parameters [$m_{\alpha}$ = 3.07(4), $m_{\beta}$ = 6.03(11), $T_D$ = 0.87(6) K,
$B_0$ = 16(8) T, $g_{\alpha}$ = 1.73(5) and $g_{\beta}$ = 1.52(12)] for all the
considered components. A very good quantitative agreement with the experimental
data is observed. Even though the data relevant to the classical orbits
$\alpha$, $\beta$ and 2$\beta$-$\alpha$ follow the Lifshitz-Kosevich formula due
to negligible second order terms, these second order terms strongly enters the
field- and temperature-dependent amplitude of the components 2$\alpha$, 2$\beta$
and alone are responsible for the 'forbidden frequency' $\beta$-$\alpha$.

\begin{acknowledgements}
The support of the European Magnetic Field Laboratory (EMFL) is acknowledged. The work was partially supported by the RFBR grant 14-03-00119-a and Program N2 of the Presidium of Russian Academy of Science.
\end{acknowledgements}


\end{document}